\documentclass[journal]{IEEEtran}

\usepackage{url, cite}
\usepackage{subfig, bm, color}
\usepackage{amsmath, graphicx, amsfonts, amssymb, algorithm, algorithmic}

\DeclareMathOperator*{\argmin}{arg\,min}

\begin{document}
\title{Dual-Functional Waveform Design with Local Sidelobe Suppression via OTFS Signaling}
\author{
   \IEEEauthorblockN{Kecheng Zhang, Weijie Yuan,\IEEEmembership{~Member,~IEEE}, Pingzhi Fan,\IEEEmembership{~Fellow,~IEEE}, Xianbin Wang,\IEEEmembership{~Fellow,~IEEE}}
   \thanks{Copyright (c) 2024 IEEE. Personal use of this material is permitted. However, permission to use this material for any other purposes must be obtained from the IEEE by sending a request to pubs-permissions@ieee.org.}
   \thanks{This work is supported in part by the National Natural Science Foundation of China under Grant 62101232, in part by Guangdong Provincial Natural Science Foundation under Grant 2022A1515011257, in part by Shenzhen Science and Technology Program under Grant JCYJ20220530114412029, and in part by the Shenzhen Key Laboratory of Robotics and Computer Vision (ZDSYS20220330160557001); the work of Pingzhi Fan was supported by NSFC project No.U23A20274.}
   \thanks{K. Zhang and W. Yuan are with the Department of Electronic and Electrical Engineering, Southern University of Science and Technology, Shenzhen 518055, China (email: zhangkc2022@mail.sustech.edu.cn; yuanwj@sustech.edu.cn).}
   \thanks{P. Fan is with the Information Coding $\&$ Transmission Key Lab of Sichuan Province, CSNMT Int. Coop. Res. Centre (MoST), Southwest Jiaotong University, Chengdu 61175, China (email: p.fan@ieee.org).}
   \thanks{X. Wang is with the Department of Electrical and Computer Engineering, Western University, London, ON N6A 5B9, Canada (e-mail: xianbin.wang@uwo.ca).}
}
\maketitle
\begin{abstract}
   Integrated sensing and communication (ISAC) is viewed as a key technology in future wireless networks. One of the main challenges in realizing ISAC is developing dual-functional waveforms that can communicate with communication receivers and perform radar sensing simultaneously. In this paper, we consider the joint design of a dual-functional orthogonal time-frequency space (OTFS) signal and a receiving filter for the ISAC system. The problem of ISAC waveform design is formulated as the minimization of the weighted integrated sidelobe level (WISL) of the ambiguity function and the interference term from ISAC waveform, with constraints on signal-to-noise ratio loss. The majorization-minimization algorithm combined with alternating iterative minimization is implemented to solve the optimization problem. Simulation results show that the WISL and the interference term can be significantly decreased to guarantee achievable data rates and detection performance.
\end{abstract}
\section{Introduction}
\label{sec:Intro}
With the rapid deployment of wireless communication technologies, many emerging applications, such as multi-functional radio frequency systems \cite{wang2014system}, and unmanned aerial vehicle communication and sensing, require both high-speed communication services and high-accuracy sensing performance simultaneously. To meet these requirements, integrated sensing and communication (ISAC) \cite{ISAC1}, which can perform sensing and communication (S$\&$C) through the same frequency band on a single platform, has been considered a promising technique.

Due to different waveform requirements for S$\&$C services, in which the communication waveform is desired to be random to convey data information while the radar waveform is expected to have an ideal autocorrelation property \cite{ISAC4}, one of the main design challenges for ISAC lies in finding the dual-functional waveform that can be employed for radar sensing and information transmission simultaneously \cite{ISAC1}. As concluded by \cite{ISAC4}, there are three main categories in ISAC waveform designs, 1) communication-centric waveform design, in which classical communication waveforms, such as orthogonal frequency-division multiplexing (OFDM), are modified to support sensing services simultaneously; 2) sensing-centric waveform design, in which communication information is embedded into classic sensing waveforms, such as the chirp-based ISAC waveform and spatial-modulated ISAC waveform; 3) joint waveform optimization, in which the ISAC waveform is designed according to performance metrics for both S$\&$C, such as SINR-based optimization, Cramér-Rao bound-based optimization. Due to the dynamics in striking S$\&$C performance tradeoffs, joint waveform optimization can be a critical challenge in ISAC, which motivates us to study it in this paper. Although there are some other concerns in ISAC waveform design, such as security and privacy guarantees within ISAC application \cite{10418473, li2022joint, yang2022secure}, we will now focus on the sidelobe suppression in the ambiguity function (AF) \cite{richards2014fundamentals, Benedetto2009} in this paper.

The AF represents the range-Doppler responses from targets with different time delays and Doppler-shifts \cite{richards2014fundamentals, Benedetto2009}. Designing the ambiguity function with low sidelobe levels, which is crucial for enhancing the resolution and accuracy of pulse compression radar systems through improved target discrimination and a decreased probability of false detections, has been well investigated in conventional radar waveform design. However, how to generate a waveform with a low AF sidelobe level while providing a high communication rate has been rarely considered in the literature on ISAC waveform design. The authors in [9] derived theoretical bounds on the amplitudes of the sampled grid in the AF, but they did not provide a waveform design scheme for the ISAC application. In \cite{yang2020dual}, the authors designed a dual-use waveform, in which the sidelobe of AF was reduced and the information symbols were embedded into the AF. This scheme leads to a low communication rate due to the limited number of embedded information symbols and requires a high signal-to-noise ratio (SNR) to have a low bit-error rate. In \cite{zhang2022integrated}, an ISAC waveform was designed based on a sparse vector coding scheme to obtain a low sidelobe level, but the sidelobe level only reduced by around 10 dB compared to the main lobe, which is not enough to avoid effects of clutter. Therefore, it is necessary to develop a method to suppress the sidelobe level as well as guarantee a relatively high communication rate.

In this paper, we jointly design the receiving filter and the transmit sequence to minimize interference terms caused by the ISAC waveform \cite{ISAC1} and weighted integrated sidelobe level (WISL). Compared to the separate design of the receiving filter or the transmit sequence, this approach provides more degrees of freedom (DoF) in local AF shaping \cite{he2011synthesizing}, leading to improved performance. Instead of optimizing the time domain waveform, we design the transmit sequence via orthogonal time-frequency space (OTFS) signaling \cite{10152009} since OTFS modulation can achieve reliable communications over multi-path channels with different delay and Doppler shifts in each path \cite{10152009}. Moreover, the simplified representation of multi-path channels in OTFS signaling will provide a smaller channel estimation error \cite{10152009}, thereby improving the S$\&$C performance in practical applications.

Specifically, we implement an alternately iterative majorization minimization (MM) method to solve the joint design problem of the receiving filter and the transmit sequence, by giving a closed-form solution in each iteration. Simulation results demonstrate that both the interference term and WISL can be significantly minimized, and the designed ISAC waveform can achieve good detection performance and simultaneously guarantee the achievable data rate (ADR).

\textit{Notations:} We use a boldface lowercase letter and a boldface capital letter to denote a vector and a matrix, respectively. $\otimes$ denotes the Kronecker product; $|\cdot|$ and $\|\cdot\|$ denote the absolute value and $l^2$ norm, respectively; $\mathcal{R}\{\cdot\}$ denotes the real part of a complex number; vec($\cdot$) represents for the vectorization of a matrix by stacking the columns of the matrix on top of each other. The superscript $(\cdot)^{\text{T}}$ and $(\cdot)^{\text{H}}$ represent the transpose and the conjugate transpose operations, respectively.
\section{System Model and Problem Formulation}
\label{sec:SysModel}
In this section, we will first present the OTFS signal model for communication and radar sensing, respectively. After that, we will formulate the optimization problems for the design of the receiving filter and the transmitted sequence.
\subsection{OTFS Signal Model}
\label{sec:OTFS_Signal_Model}
Consider the bandwidth and time duration of one OTFS frame being $M \Delta f$ and $NT$, respectively, where $\Delta f$ and $T$ are the bandwidth of each sub-carrier and the duration of one-time slot, $M$ and $N$ are the number of sub-carriers and time slots. The information bit sequence is mapped to $MN$ data symbols within a constellation set $\mathcal{A}$, and placed in a two-dimentional (2D) DD domain grids, $\{X_{\text{DD}}[i, j] | i \in \{0, \dots, M-1\}, j \in \{0, \dots, N-1\}\}$, where $X_{\text{DD}}$ is the $(i, j)$-th element of the DD domain symbol matrix $\mathbf{X}_{\text{DD}} \in \mathbb{C}^{M \times N}$, $i$ and $j$ denote Doppler and delay indices, respectively.

The DD domain symbols $\{X_{\text{DD}}[i, j]\}$ are mapped into the time-frequency (TF) domain $\{X_{\text{TF}}[i, j]\}$ through the inverse symplectic finite Fourier transform (ISFFT) \cite{10152009}, which is converted into a time domain signal $s(t)$ after implementing the Heisenberg transform \cite{10152009}. By sampling $s(t)$ at a sampling interval of $\frac{T}{M}$, we get the discrete-time representation of the OTFS transmitting signal,
\begin{equation}\label{eq:trans_vec}
   \mathbf{s} = (\mathbf{F}_{N}^{\text{H}} \otimes \mathbf{G}_{\text{tx}}) \mathbf{x}_{\text{DD}},
\end{equation}
where $\mathbf{x}_{\text{DD}} = \text{vec}(\mathbf{X}_{\text{DD}})$, $\mathbf{x}_{\text{DD}} \in \mathbb{C}^{N_{\text{x}} \times 1}$, $N_{\text{x}} = MN$, $\mathbf{G}_{\text{tx}} \in \mathbb{C}^{M\times M}$ is a diagonal matrix with the sample of $g_{\text{tx}}(t)$ at a sampling interval of $\frac{T}{M}$ as its entries, $\mathbf{s}\in \mathbb{C}^{N_{\text{x}} \times 1}$, and $\mathbf{F}_N \in \mathbb{C}^{N \times N}$ is the Fourier transform matrix, such that $\mathbf{F}_{N}^{\text{H}}\mathbf{F}_{N} = \mathbf{I}_{N}$. We consider the pulse shaping filter to be a rectangular pulse\footnote{For ease of exposition, we considered a rectangular pulse here. The proposed ISAC waveform design scheme can also be implemented for other types of pulses, such as the root-raised cosine (RRC) pulse, or the linear frequency modulation (LFM) pulse.}, i.e.,
\begin{equation}\label{eq:rec_pulse}
   g_{tx}(t) = \begin{cases}
      1, \text{ if } 0 \leq t \leq T, \\
      0, \text{ otherwise},
   \end{cases}
\end{equation}
which makes $\mathbf{G}_{\text{tx}} = \mathbf{I}_{M}$. We assume that a cyclic prefix (CP) of length $N_{\text{CP}}$ is appended to $\mathbf{s}$ before transmission. The transmitted signal sequence becomes $\tilde{\mathbf{s}} = \mathbf{\Gamma} \mathbf{s} \in \mathbb{C}^{(N_{\text{x}} + N_{\text{CP}}) \times 1}$, where $\mathbf{\Gamma}_{\text{CP}} \in \mathbb{C}^{N_{\text{CP}} \times MN}$, and
\begin{equation}
   \mathbf{\Gamma} = \left[ \mathbf{\Gamma}_{\text{CP}}^{\text{T}}, \mathbf{I}_{MN} \right]^{\text{T}}, \mathbf{\Gamma}_{\text{CP}}[i, j] = \left\{ \begin{matrix}
      1, & j = i + N_{\text{x}} - N_{\text{CP}}, \\
      0, & \text{otherwise}.
   \end{matrix} \right.\nonumber
\end{equation}
The received signal after passing a multi-path time-varying wireless channel is,
\begin{equation}\label{eq:receive_signal}
   r(t)=\iint h(\tau,\nu)e^{j2\pi\nu(t-\tau)}s(t-\tau) \,d\tau\,d\nu+z(t),
\end{equation}
where $z(t)$ denotes the additive white Gaussian noise (AWGN) process with one side power spectral density $\sigma^2$, and $h(\tau,\nu)\in \mathbb{C}$ is the complex base-band channel impulse response in the DD domain \cite{10152009}, $h(\tau,\nu)=\sum_{i=1}^{P}h_{i} \delta(\tau-\tau_{i}) \delta(\nu-\nu_{i})$, where $P\in \mathbb{Z}$ is the number of resolvable paths, and $h_{i}$, $\tau_{i} = \frac{l_{\tau_{i}}}{M\Delta f}$ and $\nu_{i} = \frac{k_{\nu_{i}}}{NT}$ denote the channel coefficient, delay, and Doppler shift associated with the $i$th path, $l_{\tau_{i}}$ and $k_{\nu_{i}}$ are the delay and Doppler indices of the $i$-th path, respectively. By sampling $r(t)$ at the sampling interval of $\frac{T}{M}$ and discarding the CP, we have the discrete representation of the received signal \cite{raviteja2018practical}
\begin{equation}\label{eq: TD_Received_Vec_Form}
   \mathbf{r} = \mathbf{H} \mathbf{s} + \mathbf{z},
\end{equation}
where $\mathbf{H} \in \mathbb{C}^{N_{\text{x}} \times N_{\text{x}}}$ is the effective channel matrix,
\begin{equation}\label{eq:Channel}
   \begin{aligned}
      \mathbf{H}     & = \sum_{i=1}^{P} h_i \mathbf{\Pi}_{l_{i}} \mathbf{D}_{k_{i}}, \mathbf{\Pi}_{l} = \left[\begin{matrix}
            \mathbf{0}_{(N_{\text{x}} - l) \times l} & \mathbf{I}_{N_{\text{x}} - l}           \\
            \mathbf{I}_{l}              & \mathbf{0}_{l\times (N_{\text{x}} - l)} \\
         \end{matrix}\right], \\
      \mathbf{D}_{k} & = \text{diag}\{\mathbf{d}_{k}\}, \mathbf{d}_{k} = [1, e^{-\frac{j2\pi k}{N_{\text{x}}}}, \dots, e^{-\frac{j2\pi(N_{\text{x}} - 1)k}{N_{\text{x}}}}]^{\text{T}}.
   \end{aligned}
\end{equation}
The designed ISAC transmit sequence (denoted as $\mathbf{x}$) is usually different from the original OTFS signal $\mathbf{s}$ in \eqref{eq:trans_vec}, which will lead to interference, denoted as $(\mathbf{H} \mathbf{x} - \mathbf{s})$, at the communication receiver. To better show the relationship between the interference caused by ISAC waveform and the received signal, the receiving signal is represented as follows, which is similar to the model considered in \cite{ISAC1},
\begin{equation}\label{eq:ISAC_TD_Model}
   \mathbf{r} = \mathbf{s} + (\mathbf{H} \mathbf{x} - \mathbf{s}) + \mathbf{z},
\end{equation}
where $\mathbf{z}$ follows the complex Gaussian distribution $\mathcal{CN}(0, \sigma^2)$, and $\sigma^2$ is the noise power.

By implementing the Wigner transform and the SFFT \cite{10152009} (the opposite transformation of the Heisenberg transformation and ISFFT, respectively), the time domain received signal $\mathbf{r}$ is converted to the DD domain signal whose vectorized discrete representation (denoted as $\mathbf{y}$) under the rectangular receiving pulse shaping filter \cite{raviteja2018practical} is
\begin{equation}\label{eq:DD_IO}
   \begin{aligned}
      \mathbf{y} & = (\mathbf{F}_{N} \otimes \mathbf{I}_{M}) \mathbf{r}                                                                                             \\
                 & = (\mathbf{F}_{N} \otimes \mathbf{I}_{M}) \mathbf{s} + (\mathbf{F}_{N} \otimes \mathbf{I}_{M}) (\mathbf{H} \mathbf{x} - \mathbf{s} + \mathbf{z}) \\
                 & = \mathbf{x}_{\text{DD}} + \tilde{\mathbf{z}}_{\text{DD}},
   \end{aligned}
\end{equation}
where $\tilde{\mathbf{z}}_{\text{DD}} = (\mathbf{F}_{N} \otimes \mathbf{I}_{M}) (\mathbf{H} \mathbf{x} - \mathbf{s} + \mathbf{z})$ is the effective DD domain noise-plus-interference term, and $\mathbb{E}[\tilde{\mathbf{z}}_{\text{DD}}^{\text{H}} \tilde{\mathbf{z}}_{\text{DD}}] = \|\mathbf{H} \mathbf{x} - \mathbf{s}\|^2 + \mathbb{E}[\|\mathbf{z}\|^2]$. It is clear that the smaller the interference term $\| \mathbf{H} \mathbf{x} - \mathbf{s} \|^2$, the better the communication performance.
\subsection{Radar Model}
\label{sec:RadarModel}
Enforcing a low sidelobe level among the designed waveform as well as requiring good communication performance significantly complicates the waveform design problem and reduces the DoF in the design procedure. It may even be possible that the corresponding optimization problem is infeasible to meet the communication and the sidelobe level requirements. To avoid the difficulty, we consider the joint design of a receiving filter as well as the transmitted sequence for ISAC.

For the transmitted signal sequence $\tilde{\mathbf{x}} = \mathbf{\Gamma} \mathbf{x} \in \mathbb{C}^{N_{\text{h}} \times 1}$ and a receiving filter $\mathbf{h} \in \mathbb{C}^{N_{\text{h}} \times 1}$, $N_{\text{h}} = N_{\text{x}} + N_{\text{cp}}$, the narrow band cross ambiguity function (CAF) \cite{Benedetto2009} between them is denoted as $f$: $\mathbb{C}^{N_{\text{h}} \times 1} \times \mathbb{C}^{N_{\text{h}} \times 1} \rightarrow \mathbb{C}$:
\begin{equation}
   f_{lk}(\mathbf{h}, \mathbf{x}) = \frac{1}{N_{\text{x}}} \sum_{n=0}^{N_{\text{x}}-1} h_{a}[n]^{*} \tilde{x}_{a}[(l+n)] e^{-\frac{j2\pi n k}{N_{\text{x}}}},
\end{equation}
where $(\cdot)_{a}$ dentoes the aperiodic extension \cite{Benedetto2009}. The CAF above can be expressed in vector form as $f_{lk}(\mathbf{x}, \mathbf{h}) = \mathbf{h}^{\text{H}} \mathbf{J}_{l} \mathbf{D}_{k} \mathbf{\Gamma} \mathbf{x}$, where $\mathbf{D}_{k}$ are defined in \eqref{eq:Channel}, and
\begin{equation}
   \mathbf{J}_{l} = \left[\begin{matrix}
         \mathbf{0}_{(N_{\text{h}} - l) \times l} & \mathbf{I}_{N_{\text{h}} - l}           \\
         \mathbf{0}_{l}                           & \mathbf{0}_{l\times (N_{\text{h}} - l)} \\
      \end{matrix}\right], \mathbf{J}_{-l} = \mathbf{J}_{l}^{\text{T}}.
\end{equation}

Instead of minimizing the ISL on the whole delay and Doppler grid, we focus on minimizing the ISL only on an interested local area due to the impossibility of constructing an unimodular sequence with a thumbtack-like ambiguity function \cite{Benedetto2009}. Denote the interested local area as $\mathbf{\Omega} = \{(l,k) | l \in \mathbf{\Omega}_{L}, k \in \mathbf{\Omega}_{K}, \omega_{lk} \neq 0\}$, where $\mathbf{\Omega}_{L}$ and $\mathbf{\Omega}_{K}$ denote the delay and Doppler indices of the interested area in the AF, respectively, we define the weighted ISL (WISL) as
\begin{equation}
   \text{WISL} = \sum_{(l,k) \in \mathbf{\Omega}} \omega_{lk} |f_{lk}(\mathbf{x},\mathbf{h})|^2,
\end{equation}
where $\omega_{lk}$ is the weight at delay-Doppler grid $(l,k)$, and $\omega_{00} = 0$ since the mainlobe level should not be suppressed.

It should be noted that, compared to the matched filter which is optimal for white noise, the mismatched filter will result in a loss in processing gain (LPG) \cite{Wang2022}, and it is defined as
\begin{equation}
   \begin{aligned}
      \text{LPG} & =10\log_{10}\left(\frac{|\mathbf{h}^{\text{H}} \tilde{\mathbf{x}}|^2}{\|\mathbf{h}\|^2 \|\tilde{\mathbf{x}}\|^2}\right) \leq 0.
   \end{aligned}
\end{equation}
The relationship between the LPG $(\beta^2)$ and the SNR loss $(\mu, \text{in dB})$ is $\beta = \sqrt[]{P_{\text{h}} N_{\text{h}}} \cdot 10^{-\mu/20}$, where $P_{\text{h}}$ is the power of receiving filter. We can control the LPG through a peak loss function $f_{\text{PL}}(\tilde{\mathbf{x}}, \mathbf{h}) = |\mathbf{h}^{\text{H}} \mathbf{\Gamma} \mathbf{x} - \beta|^2$.
\section{Joint Design of Sequence and Mismatched Filter via the Majorization Minimization Method}
\label{sec:Method}
\subsection{Problem Formulation}
\label{sec:ProbForm}
To guarantee the sensing and communication performance, we need to minimize the WISL by controlling the peak loss at the AF, which is similar to the WISL suppression problem considered in \cite{Wang2022}, and reduce the interference term at the receiver simultaneously. The corresponding optimization problem can be expressed as
\begin{equation} \label{prob:origin_prob}
   \begin{aligned}
      \min_{\mathbf{h}, \mathbf{x}} \text{ } & \rho [\text{WISL} + f_{\text{PL}}(\mathbf{x},\mathbf{h})] + (1-\rho) \|\mathbf{H} \mathbf{x} - \mathbf{s}\|^2 \\
                                             & \text{s.t. } \|\mathbf{h}\|^2_2 = P_{\text{h}}, |x_{n}|^2 = 1, n=1,\dots,N_{\text{x}},
   \end{aligned}
\end{equation}
where $\rho$ is the pareto weight between the sensing and communication terms. Denoting the objective function in \eqref{prob:origin_prob} as $g(\mathbf{x},\mathbf{h})$, we can solve the optimization problem above using a Gauss-Seidel scheme. Assuming we have iterated $(i-1)$ times, where the corresponding receiving filter and transmitting sequence are $\mathbf{h}^{(i-1)}$ and $\mathbf{x}^{(i-1)}$, respectively, the solution for the $i$-th iteration can be expressed as
\begin{align}
   \mathbf{h}^{(i)} & = \argmin_{\mathbf{h}} g(\mathbf{x}^{(i-1)},\mathbf{h}), \label{prob:solve_h} \\
   \mathbf{x}^{(i)} & = \argmin_{\mathbf{x}} g(\mathbf{x},\mathbf{h}^{(i)}). \label{prob:solve_x}
\end{align}

Due to the unimodular constraints on $\mathbf{x}$, the MM method, which is generally well-suited for non-convex and non-smooth problems, is used to solve \eqref{prob:solve_h} and \eqref{prob:solve_x}, where the majorization technique is implemented to simplify the optimization process.
\subsection{Solution for the mismatched filter}
\label{sec:MismatchedSol}
Under each $k$, denote $\mathbf{A}_{k} = \sum_{l \in \mathbf{\Omega}_{L}} \omega_{lk}^{\prime} \text{vec}(\mathbf{J}_{l}) \text{vec}(\mathbf{J}_{l})^{\text{H}}$, $\mathbf{P}_{k} = \mathbf{D}_{k} \mathbf{\Gamma} \mathbf{x} \mathbf{h}^{\text{H}}$, where $\omega_{00}^{\prime} = \rho$ and $\omega_{lk}^{\prime} = \omega_{lk}$ for all other $l$ and $k$. After some mathematical operations, problem \eqref{prob:solve_h} can be expressed as
\begin{equation}\label{prob_h_before_MM}
   \begin{aligned}
      \min_{\mathbf{h}} & \sum_{k \in \mathbf{\Omega}_{K}} \text{vec}(\mathbf{P}_{k}^{\text{H}})^{\text{H}} \mathbf{A}_{k} \text{vec}(\mathbf{P}_{k}^{\text{H}}) - 2 \beta \mathcal{R}\{tr(\mathbf{P}_{0})\} \\
                        & \text{s.t. } \|\mathbf{h}\|^2 = P_{\text{h}}.
   \end{aligned}
\end{equation}
According to Lemma 1 in \cite{Song2016}, for each $k$, we can choose $\mathbf{M}_{k} = \lambda_{a,k} \mathbf{I} \succeq \mathbf{A}_{k}$, where $\mathbf{A} \succeq \mathbf{B}$ refers to $\mathbf{A} - \mathbf{B}$ is positive semi-definite, and $\lambda_{a,k}$ denotes the maximum eigenvalue of $\mathbf{A}_{k}$, to construct the majorization function. There is a closed solution for $\lambda_{a,k}$ under each $k$. Consider a vector set $\{\text{vec}(\mathbf{J}_{l})\}_{l=1-N_{\text{h}}}^{N_{\text{h}}-1}$,
\begin{equation}
   \begin{aligned}
      \mathbf{A}_{k} \text{vec}(\mathbf{J}_{l}) & = \sum_{j \in \mathbf{\Omega}_{L}} \omega_{jk}^{\prime} \text{vec}(\mathbf{J}_{j})\text{vec}(\mathbf{J}_{j})^{\text{H}} \text{vec}(\mathbf{J}_{l}) \\
                                                & = \omega_{lk}^{\prime} (N_{\text{h}} - |l|) \text{vec}(\mathbf{J}_{l}).
   \end{aligned}
\end{equation}
Thus, $\omega_{lk}^{\prime} (N_{\text{h}} - |l|)$ is one of the eigenvalues of $\mathbf{A}_{k}$, and $\lambda_{a,k} = \max_{l \in \mathbf{\Omega}_{L}} \omega_{lk}^{\prime} (N_{\text{h}} - |l|)$. Given $\mathbf{h}^{(t)}$ at the $t$-th iteration, the quadratic term in the objective function of problem \eqref{prob_h_before_MM} can be majorized by
\begin{equation}\label{eq: MM_h}
   \begin{aligned}
       & \sum_{k \in \mathbf{\Omega}_{K}} \left[ \lambda_{a,k} \text{vec}(\mathbf{P}_{k}^{\text{H}})^{\text{H}}  \text{vec}(\mathbf{P}_{k}^{\text{H}})\right.                               \\
       & + 2\mathcal{R}\{ \text{vec}(\mathbf{P}_{k}^{\text{H}})^{\text{H}} (\mathbf{A}_{k} - \lambda_{a,k} \mathbf{I}) \text{vec}(\mathbf{P}_{k, \mathbf{h}^{(t)}}^{\text{H}}) \}             \\
       & \left.+ \text{vec}(\mathbf{P}_{k, \mathbf{h}^{(t)}}^{\text{H}})^{\text{H}} (\lambda_{a,k} \mathbf{I} - \mathbf{A}_{k}) \text{vec}(\mathbf{P}_{k, \mathbf{h}^{(t)}}^{\text{H}})\right],
   \end{aligned}
\end{equation}
where $\mathbf{P}_{k, \mathbf{h}^{(t)}} = \mathbf{D}_{k} \mathbf{\Gamma} \mathbf{x} [\mathbf{h}^{(t)}]^{\text{H}}$. For the first term in \eqref{eq: MM_h}, we have $\text{vec}(\mathbf{P}_{k}^{\text{H}})^{\text{H}}  \text{vec}(\mathbf{P}_{k}^{\text{H}}) = tr(\mathbf{P}_{k}^{\text{H}} \mathbf{P}_{k}) = tr(\mathbf{h} \mathbf{x}^{\text{H}} \mathbf{\Gamma}^{\text{H}} \mathbf{\Gamma} \mathbf{x} \mathbf{h}^{\text{H}}) = N_{\text{h}} P_{\text{h}}$, which is a constant under the constraint $\|\mathbf{h}\|^2 = P_{\text{h}}$. Meanwhile, its third term is also a constant while solving for $\mathbf{h}^{(t+1)}$.

Substitute $\mathbf{A}_{k}$ and $\mathbf{P}_{k, \mathbf{h}^{(t)}}$ back and omit the terms irrelevant about $\mathbf{h}$, the optimization problem can be simplified as
\begin{align}\label{prob_h_final}
   \min_{\mathbf{h}} & \text{ } \mathcal{R} \left\{\mathbf{h}^{\text{H}} \left[ \sum_{k \in \mathbf{\Omega}_{K}} \left(\mathbf{\Phi}_{k} \mathbf{D}_{k} \mathbf{\Gamma} \mathbf{x} - N_{\text{h}} \lambda_{a,k} \mathbf{h}^{(t)} \right) - \beta \mathbf{\Gamma} \mathbf{x} \right] \right\} \nonumber \\
                     & \text{s.t. } \|\mathbf{h}\|^2 = P_{\text{h}},
\end{align}
where $\mathbf{\Phi}_{k} = \sum_{l \in \mathbf{\Omega}_{L}} \omega_{lk}^{\prime} tr(\mathbf{P}_{k,\text{h}^{(t)}}^{\text{H}} \mathbf{J}_{l}^{\text{H}}) \mathbf{J}_{l}$. By applying the Lagrange multiplier method, we can get the closed-form solution
\begin{equation}\label{eq:h_sol}
   \begin{aligned}
       & \mathbf{h}^{(t+1)} = \sqrt[]{\frac{P_{\text{h}}}{\|\mathbf{y}_{h^{(t)}}\|^2}}\mathbf{y}_{h^{(t)}},                                                                                                      \\
       & \mathbf{y}_{h^{(t)}} = \sum_{k \in \mathbf{\Omega}_{K}} [N_{\text{h}} \lambda_{a,k} \mathbf{h}^{(t)} - \mathbf{\Phi}_{k} \mathbf{D}_{k} \mathbf{\Gamma} \mathbf{x}] + \beta \mathbf{\Gamma} \mathbf{x}.
   \end{aligned}
\end{equation}

\subsection{Solution for the transmitted sequence}
\label{sec:SequenceSol}
Let $\mathbf{B}_{k} = \sum_{l \in \mathbf{\Omega}_{L}} \omega_{lk}^{\prime} \text{vec}(\mathbf{J}_{l}^{\text{H}}) \text{vec}(\mathbf{J}_{l}^{\text{H}})^{\text{H}}$, problem \eqref{prob:solve_x} can be rewritten as
\begin{equation}\label{eq_MM_xHBx}
   \begin{aligned}
       & \min_{\mathbf{x}} \rho \left[\sum_{l \in \mathbf{\Omega}_{L}} \text{vec}(\mathbf{P}_{k})^{\text{H}} \mathbf{B}_{l} \text{vec}(\mathbf{P}_{k}) - 2 \beta \mathcal{R}\{\mathbf{P}_{0}\}\right] \\
       & + (1-\rho) \left[\mathbf{x}^{\text{H}}\mathbf{H}^{\text{H}}\mathbf{H}\mathbf{x} - 2 (1-\rho) \mathcal{R}\{\mathbf{x}^{\text{H}}\mathbf{H}^{\text{H}}\mathbf{s}\} \right]                     \\
       & \quad\quad \text{s.t. } |x_n|=1\text{ for }n=1, \dots, N_{\text{x}}.
   \end{aligned}
\end{equation}

Now we majorize the two quadratic terms in \eqref{eq_MM_xHBx} separately. Let $\lambda_{b,k} \mathbf{I} \succeq \mathbf{B}_{k}$ for each $k$, in whcih $\lambda_{b,k}$ is the largest eigenvalue of $\mathbf{B}_{k}$, and we have the closed form solution $\lambda_{b,k} = \max_{l \in \mathbf{\Omega}_{L}} \omega_{lk}^{\prime} (N_{\text{h}} - |l|)$. Meanwhile, let $\lambda_{h} \mathbf{I} \succeq \mathbf{H}^{\text{H}} \mathbf{H}$, and $\lambda_{h}$ is the largest eigenvalue of the matrix $\mathbf{H}^{\text{H}}\mathbf{H}$. Then, similarly to \eqref{eq: MM_h}, we have the majorized problem given $\mathbf{x}^{(t)}$ at the $t$-th iteration,
\begin{equation}\label{eq:MM_x}
   \begin{aligned}
       & \min_{\mathbf{x}} \sum_{k \in \mathbf{\Omega}_{K}} (1-\rho)\mathcal{R}\{\mathbf{x}^{\text{H}} \left[(\mathbf{H}^{\text{H}}\mathbf{H} - \lambda_{h}\mathbf{I}) \mathbf{x}^{(t)} - \mathbf{H}^{\text{H}}\mathbf{s}\right]\} \\
       & + \rho \left[\mathcal{R}\{\text{vec}(\mathbf{P}_{k})^{\text{H}} (\mathbf{B}_{k} - \lambda_{b,k} \mathbf{I}) \text{vec}(\mathbf{P}_{k,\mathbf{x}^{(t)}}) - \beta \mathbf{\mathbf{P}}_{0}\}\right]                          \\
       & \text{s.t. } |x_n|=1\text{ for }n=1, \dots, N_{\text{x}},
   \end{aligned}
\end{equation}
where $\mathbf{P}_{k,\mathbf{x}^{(t)}} = \mathbf{D}_{k} \mathbf{\Gamma} \mathbf{x}^{(t)} \mathbf{h}^{\text{H}}$. Substituting $\mathbf{B}_{k}$ and $\mathbf{P}_{k,\mathbf{x}^{(t)}}$ back into the problem \eqref{eq:MM_x} yields
\begin{align}\label{eq:MM_x_2}
      & \min_{\mathbf{x}} \mathcal{R}\left\{ \rho \mathbf{x}^{\text{H}} \mathbf{\Gamma}^{\text{H}} \left(\sum_{k \in \mathbf{\Omega}_{K}} \mathbf{D}_{k}^{\text{H}} \mathbf{\Xi}_{k}\mathbf{h} - P_{\text{h}}\lambda_{b,k} \mathbf{\Gamma} \mathbf{x}^{(t)} - \beta \mathbf{h}\right) \right. \nonumber \\
      & \left. + (1-\rho)\left[ \left(\mathbf{H}^{\text{H}}\mathbf{H} - \lambda_{h} \mathbf{I}\right)\mathbf{x}^{(t)} - \mathbf{H}^{\text{H}}\mathbf{s}\right]  \right\},
\end{align}
where $\mathbf{\Xi}_{k} = \sum_{l \in \mathbf{\Omega}_{L}} \omega_{lk}^{\prime} tr(\mathbf{P}_{k,\mathbf{x}^{(t)}} \mathbf{J}_{l}) \mathbf{J}_{l}^{\text{H}}$. Problem \eqref{eq:MM_x_2} can be rewritten as
\begin{equation}\label{prob:x_y}
   \begin{aligned}
       & \min_{\mathbf{x}} \|\mathbf{x} - \mathbf{y}_{\mathbf{x}^{(t)}}\|^2 \\
       & \text{s.t. } |x_n|=1\text{ for }n=1, \dots, N_{\text{x}},
   \end{aligned}
\end{equation}
where
\begin{equation}\label{eq:y_xt}
   \begin{aligned}
       & \mathbf{y}_{x^{(t)}} = (1-\rho)\left[\mathbf{H}^{\text{H}}\mathbf{s} -\left(\mathbf{H}^{\text{H}}\mathbf{H} - \lambda_{h} \mathbf{I}\right)\mathbf{x}^{(t)}\right]                                                                                        \\
       & + \rho \beta \mathbf{\Gamma}^{\text{H}} \mathbf{h} - \rho \mathbf{\Gamma}^{\text{H}} \left(\sum_{k \in \mathbf{\Omega}_{K}} \mathbf{D}_{k}^{\text{H}} \mathbf{\Xi}_{k}\mathbf{h} - P_{\text{h}} \lambda_{b,k} \mathbf{\Gamma} \mathbf{x}^{(t)}\right).
   \end{aligned}
\end{equation}
It is clear that problem \eqref{prob:x_y} has a closed form solution,
\begin{equation}\label{eq:x_sol}
   \mathbf{x}^{(t+1)} = e^{j \arg \left(\mathbf{y}_{x^{(t)}}\right)}.
\end{equation}
In \eqref{eq:h_sol} and \eqref{eq:x_sol}, the calculation of $\mathbf{\Phi}_{k} \mathbf{D}_{k} \mathbf{\Gamma} \mathbf{x}$ and $\mathbf{D}_{k}^{\text{H}} \mathbf{\Xi}_{k}\mathbf{h}$ can be implemented via FFT (IFFT) operations to improve computation efficiency. The proposed ISAC waveform design scheme with FFT (IFFT) operations is summarized in Algorithm \ref{MM_algorithm}. Readers can refer to \cite{Song2016} for the derivation details of Algorithm \ref{MM_algorithm}. Unlike what is stated in \cite{Song2016}, we consider the FFT representations for crosscorrelation rather than autocorrelation in this paper. In addition, the squared extrapolation method (SQUAREM) \cite{varadhan2008simple} can be employed to accelerate the convergence. Due to space limitations, the details of the SQUAREM scheme are omitted in this paper. Interested readers can refer to \cite{varadhan2008simple} for the detailed implementation of the SQUAREM scheme.
\floatname{algorithm}{Algorithm}
\begin{algorithm}
   \caption{MM algorithm for joint design of the mismatched filter $\mathbf{h}$ and transmit sequence $\mathbf{x}$}
   \label{MM_algorithm}
   \begin{algorithmic}[1]
     \REQUIRE Sequence length $N_{\text{h}}$, receiving filter power $P_{\text{h}}$, LPG value $\beta$, information sequence $\mathbf{s}$, CP length $N_{\text{CP}}$, effective channel matrix $\mathbf{H}$, Pareto weight $\rho$, weight coefficients $\{\omega_{lk}\} \in \mathbf{\Omega}$
     \ENSURE Local optimal solution $\mathbf{h}^{*}$ and $\mathbf{x}^{*}$
     \STATE Set $t=0$, initialize $\mathbf{h}^{(0)}$ and $\mathbf{x}^{(0)}$
     \STATE Calculate $\lambda_h$, the largest eigenvalue of $\mathbf{H}^{\text{H}}\mathbf{H}$
     \FOR{each $k \in \mathbf{\Omega}_{K}$}
     \STATE $\lambda_{a,k} = \max_{l \in \mathbf{\Omega}_{L}} \omega_{lk}^{\prime} (N_{\text{h}} - |l|)$
     \STATE $\lambda_{b,k} = \max_{l \in \mathbf{\Omega}_{L}} \omega_{lk}^{\prime} (N_{\text{h}} - |l|)$
     \STATE $\bm{\omega}_{k} = [\omega_{0,l}^{\prime}, \omega_{1,l}^{\prime}, \dots, \omega_{N_{\text{h}}-1,l}^{\prime}, 0, \omega_{N_{\text{h}}-1,l}^{\prime}, \dots, \omega_{1}^{\prime}]$
     \ENDFOR
     \REPEAT
     \FOR{each $k \in \mathbf{\Omega}_{K}$}
     \STATE $\mathbf{f}_{k} = \mathbf{F} [\mathbf{D}_{k} \mathbf{\Gamma} [\mathbf{x}^{(t)}]\text{T}, \mathbf{0}_{1 \times N_{\text{h}}}]^{\text{T}}$
     \STATE $\mathbf{r}_{k} = \frac{1}{2N} \mathbf{F}^{\text{H}} \{\mathbf{F} ([\mathbf{h}^{(t)}]^\text{T}, \mathbf{0}_{1 \times N_{\text{h}}})^{\text{T}} \circ \mathbf{f}_{k}^{*} \}$
     \STATE $\bm{\mu}_{k} = \mathbf{F} (\mathbf{r}_{k} \circ \bm{\omega}_{k})$
     \ENDFOR
     \STATE $\mathbf{y}_{h^{(t)}} = \mathbf{h}^{(t)} + \frac{\beta \mathbf{\Gamma} \mathbf{x}^{(t)} - \frac{1}{2 N_{\text{h}}}\sum_{k \in \mathbf{\Omega}_{K}} \mathbf{F}_{:, 1:N_{\text{h}}}^{\text{H}} (\bm{\mu}_{k} \circ \mathbf{f}_{k})}{N_{\text{h}} \lambda_{a,k} |\mathbf{\Omega}_{K}|}$
     \STATE $\mathbf{h}^{(t+1)} = \sqrt[]{\frac{P_{\text{h}}}{\|\mathbf{y}_{h^{(t)}}\|^2}} \mathbf{y}_{h^{(t)}}$
     \FOR{each $k \in \mathbf{\Omega}_{K}$}
     \STATE $\mathbf{g}_{k} = \mathbf{F} ([\mathbf{h}^{(t+1)}]^\text{T}, \mathbf{0}_{1 \times N_{\text{h}}})^{\text{T}}$
     \STATE $\mathbf{t}_{k} = \mathbf{F}^{\text{H}} \{ (\mathbf{F} ([\mathbf{h}^{(t+1)}]^\text{T}, \mathbf{0}_{1 \times N_{\text{h}}})^{\text{T}})^{*} \circ \mathbf{f}_{k} \}$
     \STATE $\bm{\nu}_{k} = \mathbf{F} (\mathbf{t}_{k} \circ \bm{\omega}_{k})$
     \STATE $\mathbf{\Xi}_{k}\mathbf{h} = \frac{1}{2 N_{\text{x}}} \mathbf{F}_{:, 1:N_{\text{h}}}^{\text{H}} (\bm{\nu}_{k} \circ \mathbf{g}_{k})$
     \ENDFOR
     \STATE Calculate $\mathbf{y}_{x^{(t)}}$ via \eqref{eq:y_xt}
     \STATE $\mathbf{x}^{(t+1)} = e^{j \arg (\mathbf{y}_{x^{(t)}})}$
     \STATE $t \leftarrow t+1$
     \UNTIL{convergence}
   \end{algorithmic}
 \end{algorithm}
\section{Numerical Results}
\label{sec:illust}
We conducted independent Monte Carlo simulations to demonstrate the average performance of the designed receiving filter and OTFS transmit sequence. For each OTFS frame, we set the number of time slots $N=16$ and the number of sub-carriers $M=8$ and set $N_{\text{h}}=P_{\text{h}}$. The interested local area for sidelobe suppression was $\mathbf{\Omega}=\{ (l, k) | l \in [-5, 5], k \in [-3, 3] \}$. The time-domain effective channel matrix $\mathbf{H}$ was generated randomly for each simulation, with channel coefficients following a complex Gaussian distribution $\mathcal{CN}(0, 1)$. Each symbol in the information sequence $\mathbf{s}$ was modulated via Phase Shift Keying (PSK) modulation based on randomly generated information bits. When the difference between two successive iterations satisfies $\|\mathbf{x}^{(t+1)}-\mathbf{x}^{(t)}\| / \|\mathbf{x}^{(t)}\| \leq 10^{-6}$, we stop updating $\mathbf{x}$. Similarly, the update for $\mathbf{h}$ will be stopped when $\|\mathbf{h}^{(t+1)}-\mathbf{h}^{(t)}\| / \|\mathbf{h}^{(t)}\| \leq 10^{-6}$. The maximum number of iterations is set to $10^4$ in our simulation.

To evaluate the target detection performance of the optimized ISAC waveform, we consider a toy scenario that consists of one ISAC platform, one communication user, and two point-targets for sensing. The ISAC platform sends the optimized ISAC waveform and performs target sensing based on the signal after passing the designed receiving filter. The communication user will directly demodulate the received signal to obtain the transmitted information. The Cell-Average Constant-False-Alarm-Rate (CA-CFAR) detector \cite{richards2014fundamentals} was implemented to obtain the FAR. Unless specified, the information symbols were generated by QPSK modulation, set Pareto weight $\rho=0.5$, CP length $N_{\text{cp}}=0$, and SNR loss value $\mu = 1$ dB.
\begin{figure}[htb]
   \centering
   \subfloat[The CAF between echoes and\\transmitted communication sequence.] {
    \includegraphics[width=0.475\columnwidth]{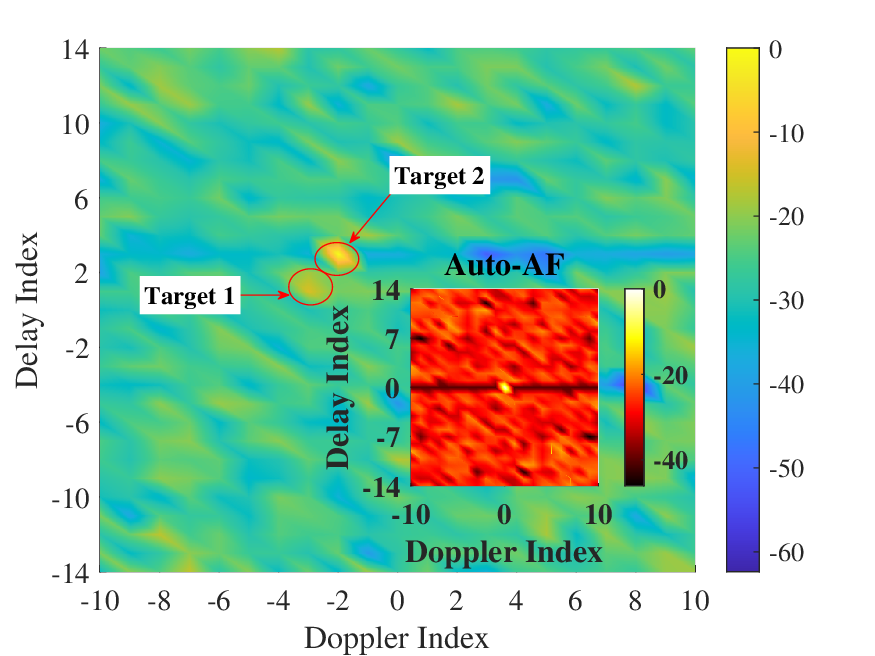}
  }
  \subfloat[The CAF optimized for ISAC\\with $\rho=0.001$.] { 
    \includegraphics[width=0.475\columnwidth]{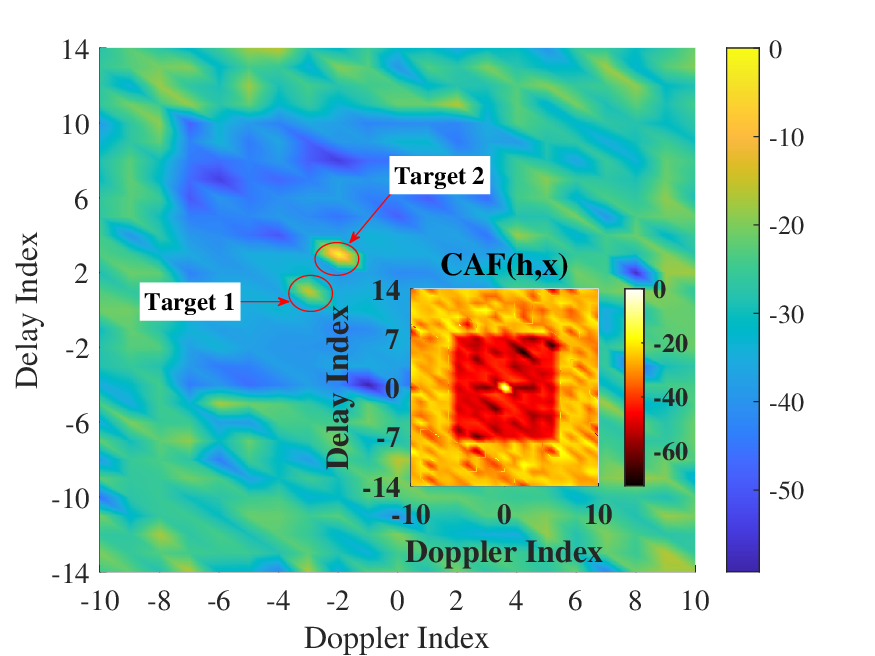}
  }
  \hfill
  \subfloat[The CAF optimized for ISAC\\with $\rho=0.5$.] {
   \includegraphics[width=0.475\columnwidth]{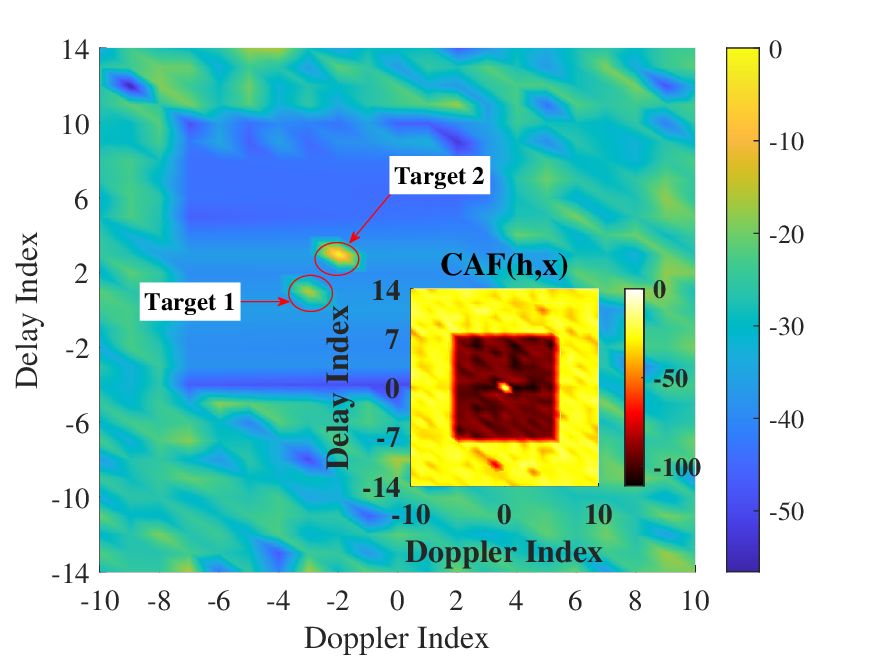}
 }
 \subfloat[The CAF optimized for ISAC\\with $\rho=1$.] {
   \includegraphics[width=0.475\columnwidth]{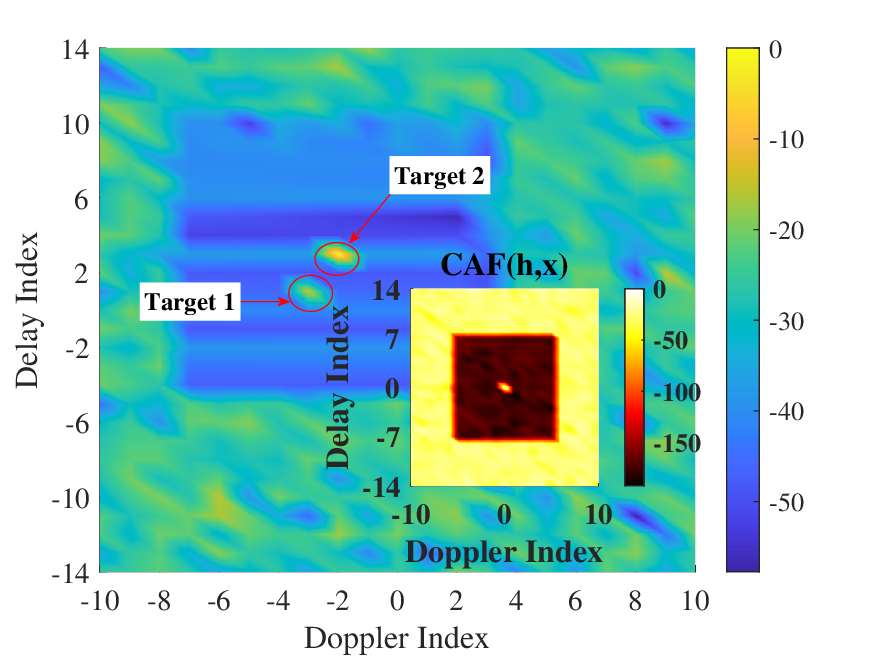}
 }
   \caption{The normalized AF and CAFs between the receiving filter and the transmit sequence under different Pareto weight $\rho$ with sequence length $N_{\text{x}}=256$ and local area $\mathbf{\Omega}=\{ (l, k) | l \in [-7, 7], k \in [-5, 5] \}$.}
   \label{CAFs}
\end{figure}

Fig. \ref{CAFs} displays the AF of the transmitted information sequence along with the CAFs optimized for different objectives, where the delay and Doppler indices for the two point targets are (1, -2), and (3, -1), respectively. The sidelobes of the CAF in Fig. \ref{CAFs}(a) are relatively high compared with the sidelobes of the CAFs shown in Fig. \ref{CAFs}(b) to (d) since the sidelobes in Fig. \ref{CAFs}(a) have amplitudes around $-20$ dB while the sidelobes in Fig. \ref{CAFs}(b) to (d) range from $-35$ dB to $-50$ dB. Moreover, the amplitude of the response from target 1 is $-14.2$ dB in Fig. \ref{CAFs}(a), which is very close to the sidelobe level, making it hard to discriminate target 1's response from the clutters. We observe in Fig. \ref{CAFs} that the sidelobes of the auto-AF exhibit amplitudes around $-20$ dB. Following the joint optimization of the receiving filter and transmitting waveform with a chosen parameter $\rho=0.001$, the sidelobe level experiences a notable reduction to approximately $-40$ dB, representing a significant $20$ dB decrease. Upon incrementing the Pareto weight $\rho$ from $0.001$ to $1$, as depicted in Fig. \ref{CAFs}(c) and Fig. \ref{CAFs}(d), a more pronounced reduction in the sidelobe level becomes evident. Specifically, the sidelobe levels are approximately $-90$ dB when $\rho = 0.5$ and further decrease to $-150$ dB when $\rho = 1$.
\begin{figure}[htb]
   \centering
   \subfloat[The average ADR w.r.t SNR.] {
    \includegraphics[width=0.7\columnwidth]{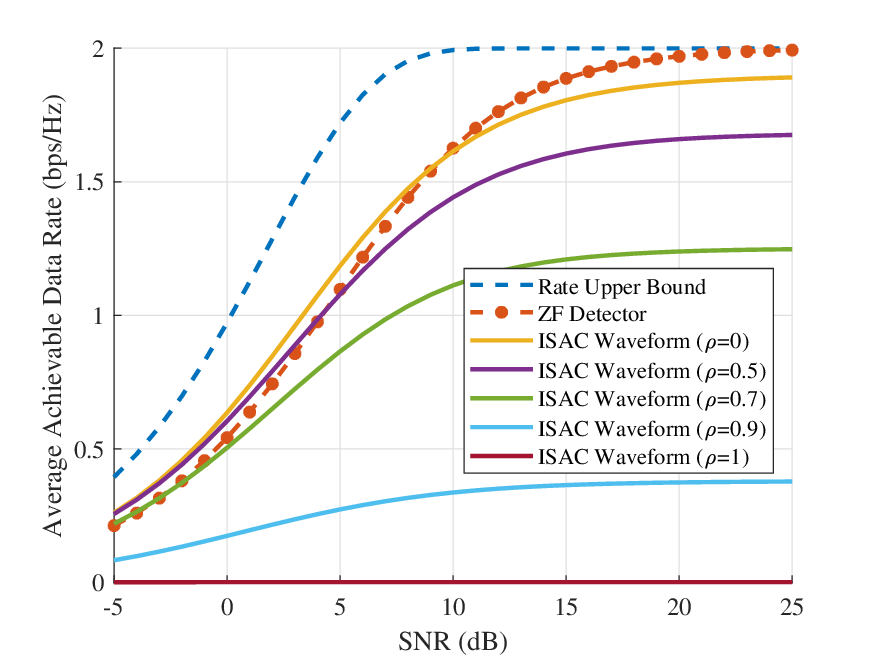}
  }
  \hfill
  \subfloat[The target detection performance w.r.t SNR.] {
    \includegraphics[width=0.7\columnwidth]{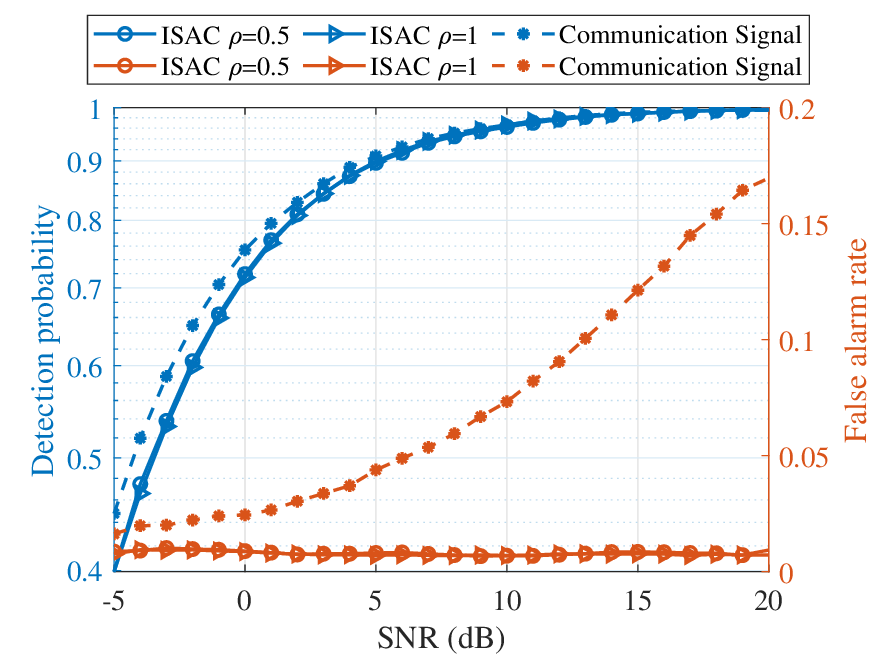}
  }
   \caption{The sensing and communication performance of the designed ISAC waveform.}
   \label{performance}
\end{figure}

Fig. \ref{performance}(a) illustrates the average ADR, calculated using mutual information between the transmitted and received symbols, under different weight values $\rho$. The average ADR becomes lower when we increase $\rho$, which means the communication performance becomes worse when we try to reduce more WISL. The optimized ISAC waveform showcases much lower FARs due to the minimized WISLs, as shown in Fig. \ref{performance}(b) while maintaining similar detection probabilities across different weight $\rho$. Observing the trade-off between WISL and interference term under different weights $\rho$, we have both good sensing and communication performance by setting $\rho=0.5$.
\begin{figure}[htb]
   \centering
   \includegraphics[width=0.995\columnwidth]{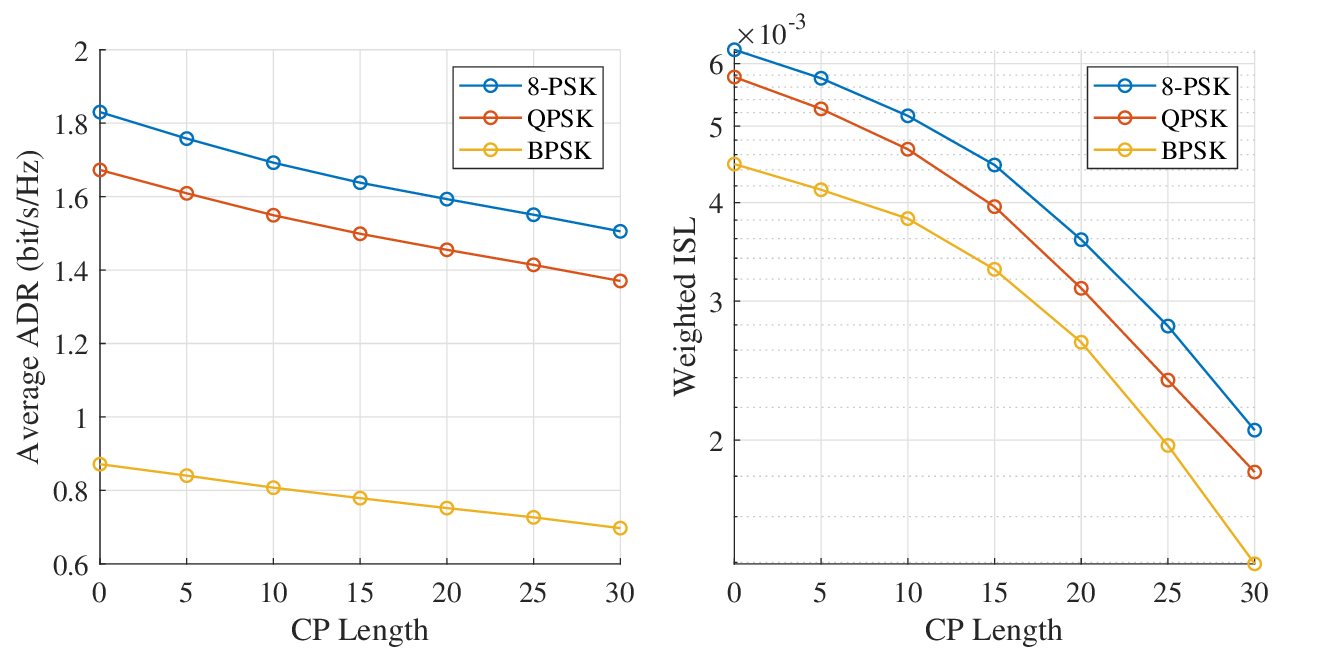}
   \caption{The average ADR and the WISL of the designed ISAC waveform with different numbers of CPs added in the transmitted sequence.}
   \label{performance_Ncp}
\end{figure}

Fig. \ref{performance_Ncp} shows the impact of the length of CP on the average ADR and WISL. The WISLs were reduced to a lower level if we added more CP (up to 25$\%$ of the sequence length) to the time domain transmitted sequence. That's because there are more DoFs to optimize the receiving filter for sidelobe-level suppression. However, with a nonzero CP, we incur a rate penalty factor $N_x/(N_x+N_{\text{CP}})$ on the effective ADR. Both the average ADR and the WISL are larger under higher modulation order. That's because minimizing the interference term under a higher modulation order is more difficult, which makes the optimization algorithm iterate more to minimize the interference term rather than the WISL. This will cause a higher WISL under a higher modulation order.
\begin{figure}[htb]
   \centering
    \includegraphics[width=0.995\columnwidth]{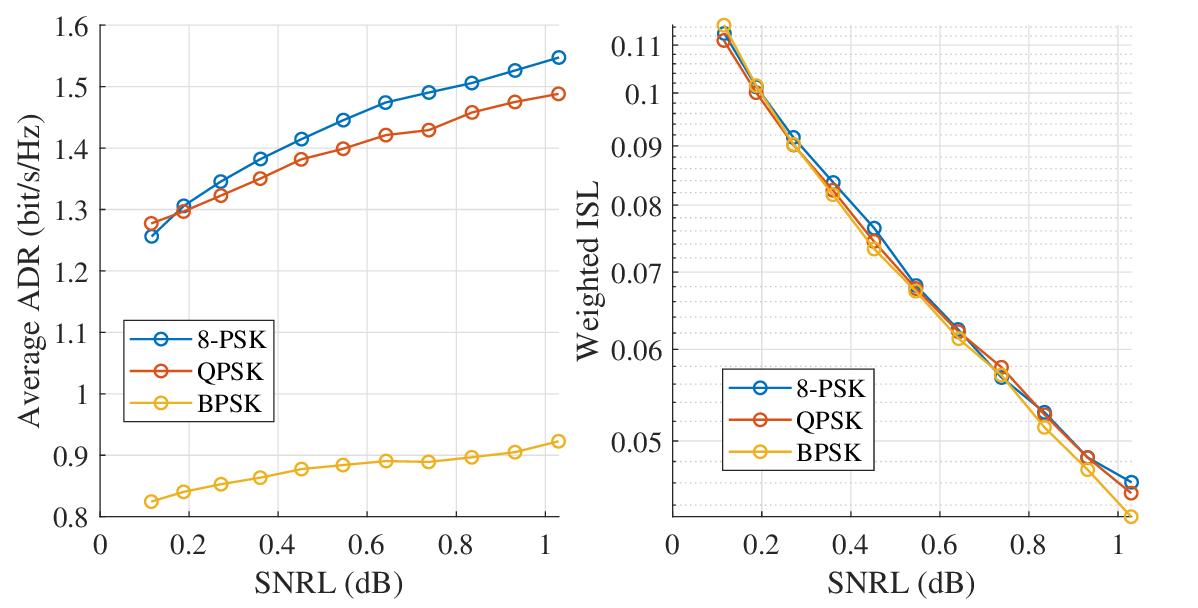}
   \caption{The average ADR and the WISL of the designed ISAC waveform under different SNRL.}
   \label{performance_SNRL}
\end{figure}

Fig. \ref{performance_SNRL} displays the average ADR and WISL under different SNR loss (or the LPG) after passing the radar receiving filter. The larger the preset SNR loss, the higher the average ADR and the smaller the WISL. That's because when we allow larger SNR loss at the radar receiver, the inner product between $\mathbf{h}$ and $\mathbf{\Gamma} \mathbf{x}$ can be smaller, which means these two vectors do not have to be closely correlated, providing more DoF to minimize the objective function. However, a larger SNR loss will lead to low SNR at the main lobe, which will reduce the detection probability. The trade-off between the SNR loss and the value of WISL, the average ADR, and the detection probability should be considered when we design the ISAC waveform.
\section{Conclusion}
\label{sec:foot}
We have introduced a joint design scheme for optimizing both the receiving filter and the OTFS transmit sequence to minimize the WISL and the interference term. The corresponding optimization problem is tackled through the MM algorithm. The numerical results demonstrate that the optimized ISAC waveform exhibits a favorable average ADR and significantly reduced FAR when compared to a signal designed solely for communication purposes. The WISL will become smaller if we add longer CP to the OTFS frame, but longer CP will reduce the effective ADR. A larger SNR loss will increase the average ADR and reduce the WISL, but the amplitude of the main lobe will become smaller, leading to low SNR at the main lobe.

\bibliographystyle{IEEEtran}
\bibliography{refs}
\vfill
\end{document}